\documentclass[useAMS,usenatbib,usegraphicx]{mn2e}

\usepackage{graphicx}
\DeclareGraphicsExtensions{.pdf,.png,.jpg,.ps}
\usepackage{epsfig}

\usepackage[bottom]{footmisc}
\usepackage{color,amsmath,enumitem}
\usepackage{threeparttable}
\usepackage{graphicx}
\DeclareGraphicsExtensions{.pdf,.eps,.jpg,.ps,.png}
\usepackage{epsfig}
\usepackage{url}
\usepackage{ amssymb }
\usepackage{pdflscape}
\usepackage{afterpage}
\usepackage{capt-of}
\usepackage{gensymb}

\title[HB Morphology of LMC GCs]{Exploring the nature and synchronicity of early cluster formation in the Large Magellanic Cloud: III. Horizontal Branch Morphology}
\author[Wagner-Kaiser et al.] {R.~Wagner-Kaiser$^1$, Dougal~Mackey$^2$, Ata~Sarajedini$^{1, 3}$, Roger~E.~Cohen$^{4}$,
\newauthor{Doug Geisler$^{5}$, Soung-Chul~Yang$^{6}$, Aaron~J.~Grocholski$^{7}$, Jeffrey~D.~Cummings$^{8}$}
\\
    $^1$University of Florida, Department of Astronomy, 211 Bryant Space Science Center, Gainesville, FL, 32611 USA\\
    $^2$Australian National University, Research School of Astronomy \& Astrophysics, Canberra, ACT 2611, Australia\\
    $^3$Florida Atlantic University, Department of Physics, 777 Glades Rd, Boca Raton, FL, 33431 USA\\
    $^4$Space Telescope Science Institute, Baltimore, MD 21218, USA\\ 
    $^5$Departamento de Astronom\'ia, Universidad de Concepcion, Casilla 160-C, Concepcion, Chile Center for Astrophysical Sciences \\
    $^6$Korea Astronomy and Space Science Institute (KASI), Daejeon 305-348, Korea \\ 
    $^7$Department of Physics and Astronomy, Swarthmore College, Swarthmore, PA 19081, USA\\ 
    $^8$Center for Astrophysical Sciences, Johns Hopkins University, Baltimore MD 21218\\}

\begin{document}

\date{}

\pagerange{\pageref{firstpage}--\pageref{lastpage}} \pubyear{2002}

\maketitle

\label{firstpage}


\begin{abstract}
We leverage new high-quality data from Hubble Space Telescope program GO-14164 to explore the variation in horizontal branch morphology among globular clusters in the Large Magellanic Cloud (LMC). Our new observations lead to photometry with a precision commensurate with that available for the Galactic globular cluster population. Our analysis indicates that, once metallicity is accounted for, clusters in the LMC largely share similar horizontal branch morphologies regardless of their location within the system. Furthermore, the LMC clusters possess, on average, slightly redder morphologies than most of the inner halo Galactic population; we find, instead, that their characteristics tend to be more similar to those exhibited by clusters in the outer Galactic halo. Our results are consistent with previous studies showing a correlation between horizontal branch morphology and age.
\end{abstract}



\section{Introduction}

The variation in horizontal branch (HB) morphology among the Galactic system of globular clusters is known to be strongly - but not entirely - determined by metallicity. Early studies found a clear distinction between the metal-rich GCs, which generally have very red HBs, and the metal-poor GCs, with HBs that are largely populated on the blue side of the RR Lyrae instability strip (\citealt{Arp:1952,Sandage:1953}). However, this trend is not absolute and there are a number of exceptions, particularly in the intermediate-metallicity range of the Milky Way globular clusters. An early example was found by \cite{Sandage:1960} in studying several GCs (e.g.: M13, M22) that have bluer than expected HB morphologies despite their intermediate metallicities. Essentially, metallicity is not alone sufficient to explain HB morphology and additional factors are necessary; this is more succinctly referred to as the ``second parameter effect" (\citealt{Sandage:1960, van-den-Bergh:1965}).

Early suggestions to explain the second parameter effect in HB morphology were cluster-to-cluster variations in age and/or helium abundance (\citealt{van-den-Bergh:1965, van-den-Bergh:1967}); these suggestions, among others, continue to be investigated in the present era. In addition to age and helium, newer work also investigates central densities of clusters, extended blue HB tails, cluster magnitude (e.g.: mass), among other parameters (\citealt{Sarajedini:1989,Chaboyer:1992,Sarajedini:1995,Chaboyer:1996,Rosenberg:1999,Recio-Blanco:2006,Dotter:2010}; see \citealt{Catelan:2009} for a comprehensive review of proposed second parameters). Some studies have suggested that more than one additional parameter may be necessary (\citealt{Richer:1996,Milone:2014}).

While the underlying cause of the second parameter effect remains a topic of research, this phenomenon has played a significant role in our interpretation of Galactic formation. \cite{Searle:1978} found a clear demarcation between inner and outer GCs in the Galaxy, with clusters inside 8 kpc of the Galactic center dominated by blue HB morphologies at given metalliciity; beyond 8 kpc clusters with redder HB morphologies become more common (\citealt{Searle:1978, Lee:1990, Sarajedini:1999, Mackey:2005, Recio-Blanco:2006, Catelan:2009, Dotter:2010}, among others). This observation is often taken as evidence that the inner halo of the Milky Way was formed quickly and early, while the outer halo continued to develop slowly over time via accretion from dwarf or satellite galaxies.

Under such a Galactic formation model, it is expected that the younger, accreted halo clusters should share broad characteristics (luminosities, ages, abundances, etc.) with the globular clusters in satellite galaxies, such as the LMC (\citealt{Zinn:1980, Suntzeff:1992, Zinn:1993, da-Costa:2003}). Previous work has examined the HB morphology in LMC clusters (\citealt{Zinn:1993, Johnson:1999, Mackey:2004, Mackey:2004b}), largely using the metric (B - R)/(B + V + R) where B is the number of blue HB stars, V is the number of RR Lyrae variables, and R is the number of red HB stars (\citealt{Lee:1989,Lee:1994}). Early studies found a few LMC globular clusters with particularly blue HB morphology - specifically NGC 2005 (\citealt{Olsen:1998}) and Hodge 11 (\citealt{Walker:1993}) - while also showing that LMC clusters were largely co-located with the young halo clusters in the Milky Way in metallicity-HB morphology space. In addition, the analysis by Mackey \& Gilmore (2004a,b) found NGC 1916, NGC 1928, and NGC 1939 to also have very blue HBs.

More recently, the rise of the multiple population problem in globular clusters has shed new light on our understanding of the horizontal branch morphology. In the Milky Way, several studies have shown that the chemical differences among stars on the horizontal branch can lead to variations in horizontal branch morphology (\citealt{Marino:2011, Villanova:2009, Marino:2013, Gratton:2011, Milone:2014}). These studies suggest that the stars at different ends of the sodium-oxygen anti-correlation have different helium abundances, which in turn affects the extension of the HB. \cite{Milone:2014} used two different parameters to quantify HB morphology -- L1, the difference in colour between the red-giant branch and the red end of the HB, and L2, the overall colour extension of the HB. They showed that L1 correlates largely with inter-cluster variations in age, while L2 correlates most strongly with intra-cluster variations in helium abundance (which are closely linked to other internal elemental abundance variations).

In studying the HB morphology of the Galactic cluster population, \cite{Dotter:2010} used another alternative approach to quantifying the HB morphology, rather than simply counting stars in and around the instability strip. Following the logic from \cite{Sarajedini:1999}, their HB morphology measurement compares the median colour of the HB stars to the median colour of the RGB at the level of the HB (\citealt{Dotter:2010}). This metric, referred to as $\Delta$(V--I), is greater for clusters with bluer HBs and correlates well with other morphological estimates. However, this approach is less dependent on observational restrictions or choice of the instability strip boundaries in addition to possessing greater sensitivity in extreme morphological cases.

We utilise this methodology to conduct a direct comparison between the HB morphologies of globular clusters in the LMC and those in the Milky Way. It allows us to extend the HB analysis from \cite{Dotter:2010} to the LMC clusters and provide a baseline measurement of HB morphology. While another option would have been to adopt the methodology outlined by \cite{Milone:2014}, unfortunately the dearth of high-resolution spectroscopic studies of LMC clusters, in conjunction with less well-determined photometry than available for Galactic globular clusters (particularly in the UV), leaves us with little clear and consistent independent information about the chemistry of their multiple populations. Moreover, the generally lower quality photometry available for clusters in the LMC, means that measuring the two parameters defined by \cite{Milone:2014}, in particular the L1 parameter, is much more prone to significant error than measuring the median HB colour.

With deep observations of six LMC globular clusters, and supplemental observations of five additional clusters in the LMC bar region, we can examine which trends, if any, seen in the horizontal branch behavior of Milky Way clusters extend to our neighbouring satellite. Incorporating information on the HB morphologies of clusters in galaxies external to the Milky Way will help shed more light on the use of globular clusters as tracers of galactic formation and evolution.

In Section \ref{Data}, we discuss the datasets from HST. We examine HB morphology in LMC clusters and compare to Galactic GCs in Section \ref{HBmorph}. In Section \ref{conclusions}, we discuss and conclude.


\section{Data}\label{Data}

\subsection{Photometry}

The data for the six outer LMC clusters come from HST Cycle 23 program GO-14164 (PI: Sarajedini). This program obtained deep imaging in the F606W and F814W filters with the Advanced Camera for Surveys (ACS) Wide Field Channel (WFC) on HST for the clusters NGC 1466, NGC 1841, NGC 2210, NGC 2257, Hodge 11, and Reticulum. Paper I in this series (\citealt{Mackey:2017}) provides a full description of the data acquisition and process of photometric analysis and evaluation. In short, each of the six clusters was observed in the F606W filter for two orbits and F814W for three orbits. Two images in each filter were short exposure images ($\approx 50-70$s per frame), with the rest being longer exposures ($\sim 350-520$s per frame). The {\sc dolphot} software package (\citealt{Dolphin:2000}) was used to photometer the short and long exposure image sets separately. These two catalogs were quality-filtered and merged to generate the final photometric catalog. The signal-to-noise ratio of stars in the region of the horizontal branch are largely $\sim 1000$ and greater, and near the main sequence turn-off point (MSTOP) the signal-to-noise ratio is $\sim 300$. The photometric depth reliably reaches down to more than 4 magnitudes below the MSTOP. Where necessary in this work, the F606W and F814W magnitudes have been converted to V and I magnitudes through the transformations provided by \cite{Sirianni:2005}.

Photometry for the other five LMC clusters analyzed here comes from \cite{Olsen:1998}. We use their published photometric catalogs for the LMC bar clusters NGC 1754, NGC 1835, NGC 1916, NGC 2005 and NGC 2019\footnote{NGC 1898 was also included in the study by \cite{Olsen:1998} but photometry for this cluster is not included in the online repository. The photometry for NGC 1928 and 1939 described by Mackey et al. (2004b) suffers from extremely heavy field contamination as well as differential reddening, such that the HB measurements are largely uninformative.}. These clusters were observed in the F555W and F814W filters with the WPFC2 Planetary Camera with both short (40 sec and 60 sec combined, F555W and F814W respectively) and long (1500 and 1800 sec combined, F555W and F814W respectively) exposures. The images were photometered with DoPHOT (\citealt{Schechter:1993}) and the process included cosmic ray rejection and a CTE correction. As the LMC bar clusters are in high-density stellar fields, field star subtraction was modeled in detail through extensive artificial star tests. Using these tests, \cite{Olsen:1998} removed field stars from their photometry and we use these cleaned datasets for our own analysis. Crowding was found to strongly affect completeness and introduce additional photometric uncertainty. The published photometric catalogs present photometry that the authors have transformed to the V and I filters in the Johnson-Cousins system using \cite{Holtzman:1995} (equation 9). Further details on the process may be found in \cite{Olsen:1998}. Despite the removal of field stars, photometry for the bar clusters remains substantially noisier than for the outer clusters; however, the median color measurements we make to analyze horizontal branch morphology are largely robust to outliers, as discussed further in Section \ref{HBmorph}.

\begin{figure}
  \centering
    \includegraphics[width=0.5\textwidth]{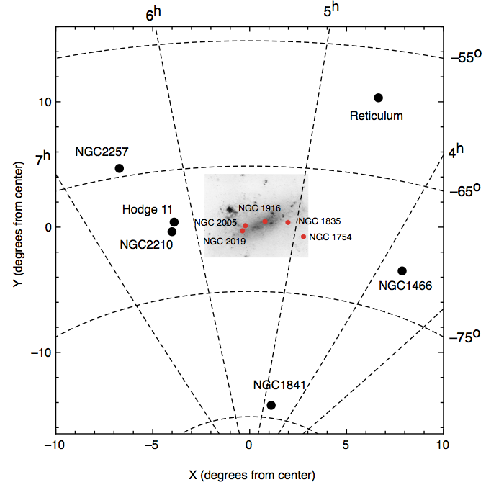}
  \caption{The LMC and surrounding region from a SkyView Digital Sky Survey composite. The six clusters from HST Cycle 23 program GO-14164 (PI: Sarajedini) are indicated in black and the five LMC bar clusters from \protect\cite{Olsen:1998} are indicated in red.}
  \label{fig:Cluster_locations}
\end{figure}

For reference, the locations of the six outer LMC clusters (Paper I) and the five bar clusters (\citealt{Olsen:1998}) are shown in Figure \ref{fig:Cluster_locations}.

\section{Horizontal Branch Morphology}\label{HBmorph}

\subsection{Measurement}\label{Measurement}

The quantification of HB morphology takes a variety of forms, from the HB type (\citealt{Dickens:1972}) to the widely used HB ratio ((B - R)/(B + V + R), \citealt{Lee:1989,Lee:1994}), and the L1 and L2 parameters defined by \cite{Milone:2014}. We elect to employ the measurement of $\Delta$(V--I), the difference in the median colour of the HB stars and the RGB stars at the level of the HB, generally following the method of \cite{Dotter:2010}. This method has a few advantages over measurement of the HB ratio (though as demonstrated by \citealt{Dotter:2010}, they remain correlated). Specifically, at the extreme ends of very red and very blue HBs, the $\Delta$(V--I) metric is more sensitive, allowing for greater differentiation between clusters. Further, the choice of the instability strip boundaries may bias the final HB ratio, but it does not influence the $\Delta$(V--I) value. The measurement of $\Delta$(V--I) is also less affected by small number statistics or observational limitations that could otherwise bias the derivation of an HB ratio. However, we do note that $\Delta$(V--I) is not as sensitive to the effects of internal helium variations on the morphology of the HB as, say, the L2 parameter introduced by \cite{Milone:2014}. Although this means that we do not have significant leverage on the effect of helium on the HB in this analysis, using Delta(V-I) we are able to provide a robust baseline comparison to the \cite{Dotter:2010} study.

In order to adequately measure the $\Delta$(V--I) metric for the LMC clusters, two measurements must be made - the median colour of the RGB at the level of the HB and the median colour of the HB. For the former, the magnitude of the HB must be determined, and for the latter it is necessary to pick the HB stars out from the CMD. To determine these values, we use the Galactic cluster NGC 5904 as a reference, as in \cite{Dotter:2010}, with photometry from the ACS Globular Cluster Treasury Program (\citealt{Sarajedini:2007}). NGC 5904 was chosen for its broad HB, allowing it to be compared to diverse cluster HB morphologies. While the chemistry of a cluster is expected to cause variations in the morphology of the HB, the overall \emph{shape} of the HB is largely consistent for different cluster morphologies (\citealt{Brown:2016,Denissenkov:2017}).

\begin{figure}
  \centering
    \includegraphics[width=0.5\textwidth]{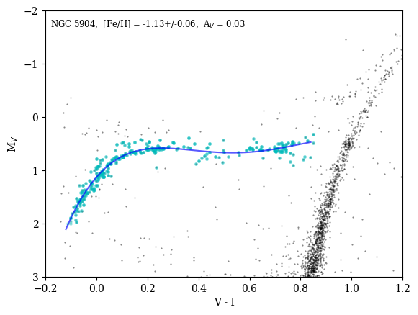}
  \caption{ACS Treasury Program photometry of NGC 5904 in black (\protect\citealt{Sarajedini:2007}). Horizontal branch stars included in the fiducial fit are shown in cyan, with the HB fiducial demarcated by the solid blue line. NGC 5904 is used in \protect\cite{Dotter:2010} and herein as a reference cluster.}
  \label{fig:5904}
\end{figure}

First, an HB fiducial for NGC 5904 is derived from the photometry. The initial fiducial estimate is made by eye at several points along the HB across the visible colour range of the HB, then fit with a radial basis function spline. To improve on the fiducial, we include stars within 2-$\sigma$ of the initial fiducial estimate and determine a moving average with bins of 0.02 in colour with a window of 0.05. A new spline is fit to these points to determine the HB fiducial. Figure \ref{fig:5904} shows the fiducial on the de-reddened, distance corrected CMD of NGC 5904.

This process is repeated to derive HB fiducials for the LMC clusters. The HB fiducial for NGC 5904 is shifted via least squares to match the HB fiducials of the LMC clusters; the difference in magnitude between the two fiducials provides the HB level for each cluster relative to NGC 5904. Using RGB stars within $\pm$0.5 magnitudes of this level, the RGB median colour is determined. The median HB colour is determined from the stars used to derive the final HB fiducial. Subtracting these two values gives us $\Delta$(V--I) for each cluster. One benefit of this approach is that the median colour is quite robust to low levels of missing stars or wrongly-included field stars. These cases will not cause a huge shift in the measured median colors except for those clusters with very few members (e.g., Reticulum).

\begin{figure*}
  \centering
    \includegraphics[width=\textwidth]{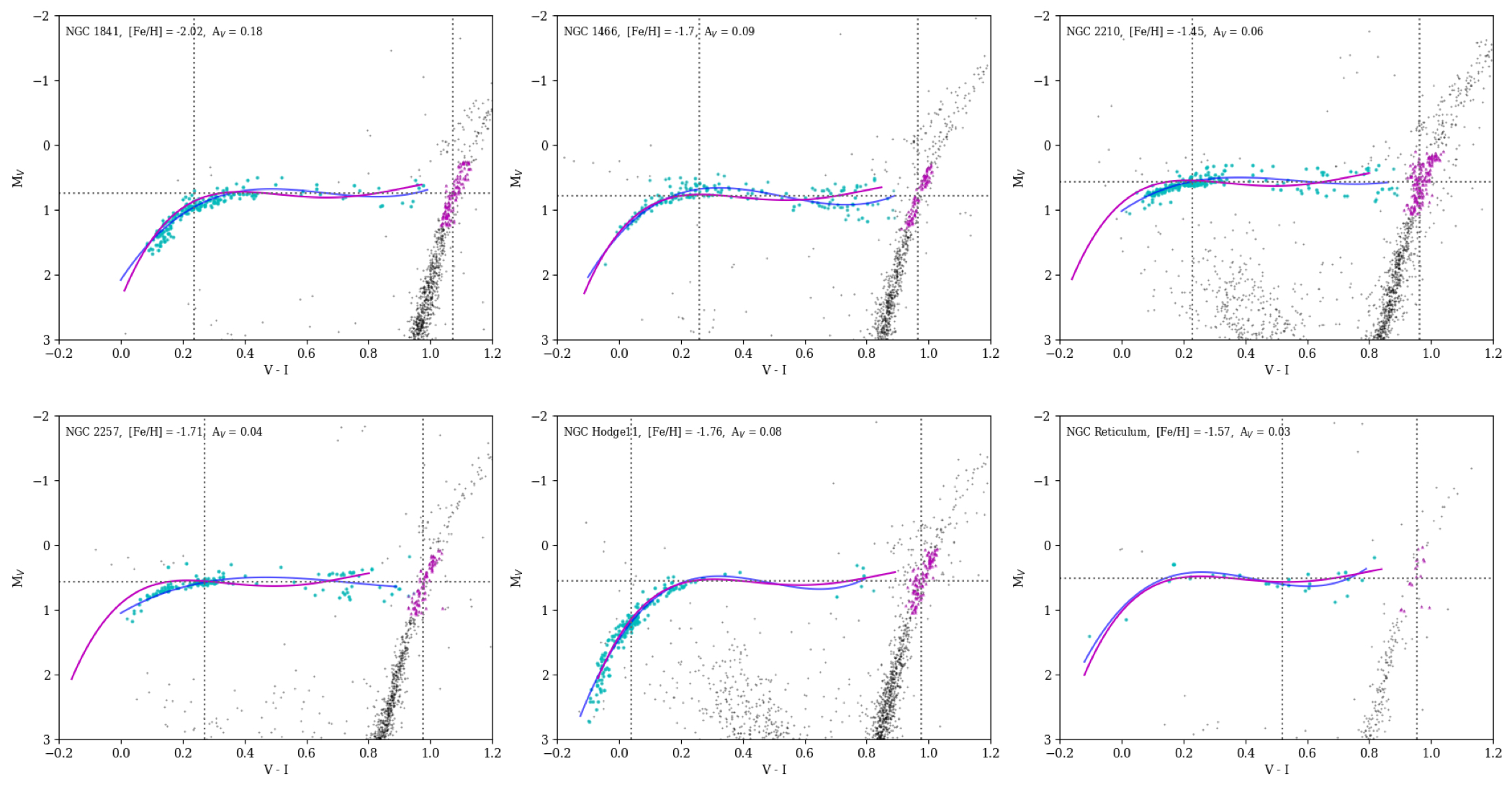}
  \caption{The $\Delta$(V--I) determination of the six outer LMC clusters from HST Cycle 23 program GO-14164. The NGC 5904 HB fiducial is indicated by the solid magenta line and the HB fiducial for the cluster in each panel is shown as the solid blue line. HB stars included in the median colour determination are shown as cyan circles in each panel; the RGB stars are indicated by magenta triangles. The horizontal line indicates the HB level of the cluster and the two vertical lines mark the median colours of the HB and RGB.}
  \label{fig:HBlmc}
\end{figure*}

\begin{figure*}
  \centering
    \includegraphics[width=\textwidth]{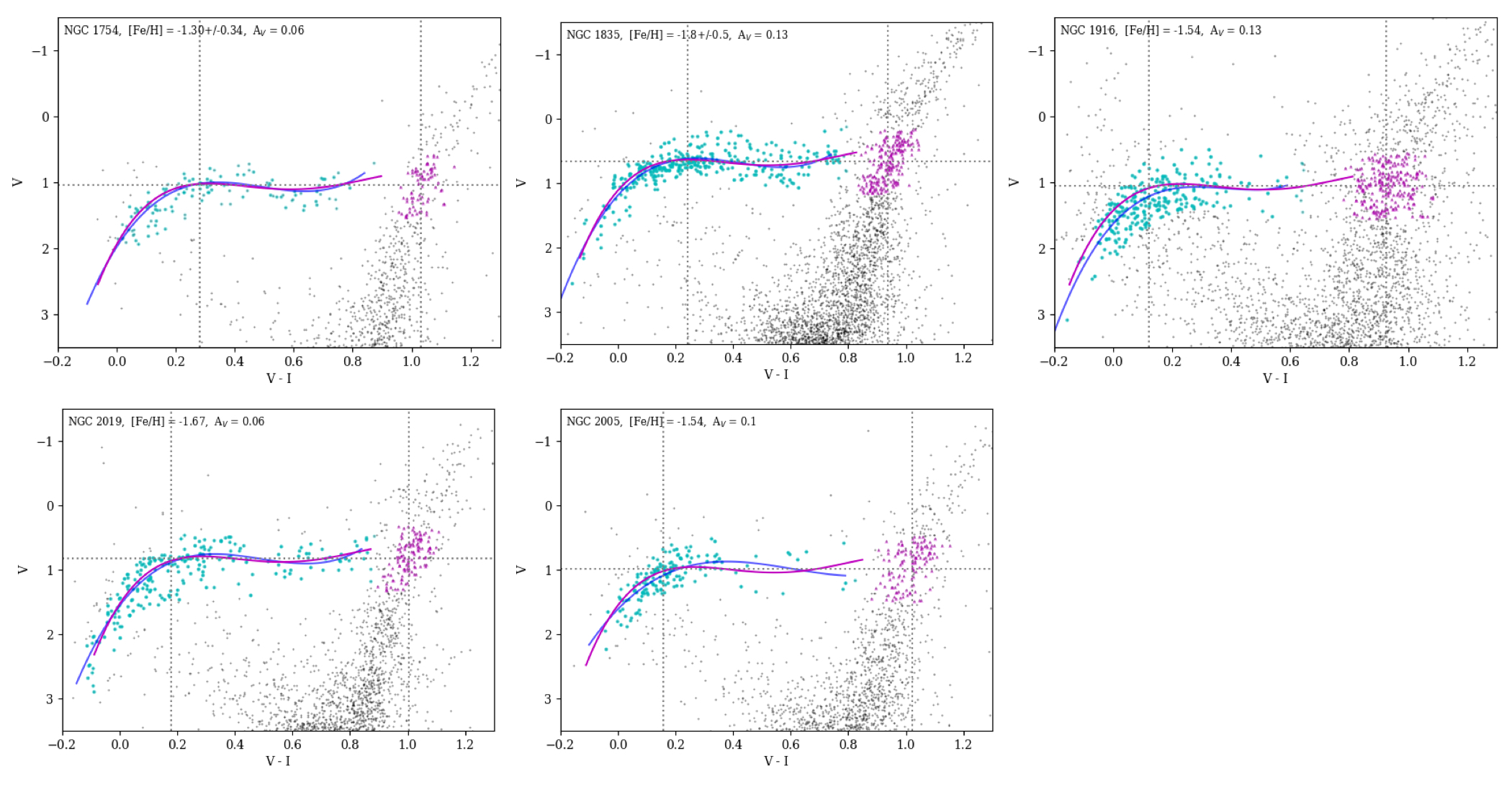}
  \caption{Same as Figure \ref{fig:HBlmc} but for the five LMC bar clusters from \protect\cite{Olsen:1998}.}
  \label{fig:HBbars}
\end{figure*}

The results of this approach are shown in Figure \ref{fig:HBlmc} for the outer LMC clusters and Figure \ref{fig:HBbars} for the LMC bar clusters, where the CMDs are de-reddened and a distance modulus of 18.5 is assumed. In these figures, the HB fiducials of each cluster are compared to NGC 5904, the HB stars and RGB stars are indicated by cyan circles and magenta triangles, respectively. The determinations of the RGB and HB median colours are also marked in each panel as vertical lines.

We present the basic properties of the LMC outer clusters from Paper I in Table \ref{tab:LMC1} and of the LMC bar clusters of the \cite{Olsen:1998} sample in Table \ref{tab:LMC2}. The derived $\Delta$(V--I) values are included in the final column with their standard error. The inherent uncertainty in the determination of $\Delta$(V--I) is larger for the Olsen et al. (1998) photometry due to greater star-to-star scatter. It is therefore possible that the results for the bar clusters could be biased relative to those measured for the outer LMC clusters of Mackey et al. (2017). This bias is difficult to quantify without any overlap in the cluster samples; however, the results for the bar clusters are broadly in line with those from previous studies using other methods of measuring horizontal branch morphology. It is also worth noting that \cite{Olsen:1998} statistically subtracted the contaminating field populations from their cluster CMDs. This effect could also, in principle, lead to a mild systematic offset between measurements for the bar clusters and those in the outer LMC. However, we expect the effect to be minimal on the overall results, as the median measurements of color are robust to outliers.

\begin{table*}
\caption{Fundamental properties of the LMC outer clusters from Paper I}
\centering
\begin{threeparttable}[b]
    \begin{tabular}{@{}|l|c|c|c|c|c|c|@{}}
\textbf{Cluster} & \textbf{[Fe/H]$_{CG97}$$^1$} & \textbf{E(B--V)$^2$}  & \textbf{Age (Gyr)$^{3}$} & \textbf{$\Delta$(V-I)} & \textbf{M$_V$$^{4}$} & \textbf{$\rho_0$$^{5}$}    \\  
\hline
NGC 1466 	&  -1.7 	& 0.09	& 13.38 $^{+ 1.67 }_{ -2.28 }$	& 0.706 $\pm$ 0.020	& 11.59	& 2.78	\\
NGC 1841 	&  -2.02 	& 0.18	& 13.77 $^{+ 1.05 }_{ -2.41 }$	& 0.836 $\pm$ 0.014	& 11.43	& 1.29	\\
NGC 2210 	&  -1.45 	& 0.06	& 11.63 $^{+ 1.80 }_{ -1.12 }$	& 0.734 $\pm$ 0.015	& 10.94	& 3.34	\\
NGC 2257 	&  -1.71 	& 0.04	& 12.74 $^{+ 1.87 }_{ -2.18 }$	& 0.704 $\pm$ 0.021	& 12.62	& 1.73	\\
Hodge11 	&  -1.76 	& 0.08	& 13.92 $^{+ 1.48 }_{ -2.01 }$	& 0.939 $\pm$ 0.011	& 11.93	& 2.59	\\
Reticulum 	&  -1.57 	& 0.03	& 13.09 $^{+ 2.21 }_{ -1.98 }$	& 0.434 $\pm$ 0.047	& 14.25	& N/A	\\
\hline
    \end{tabular}
    \begin{tablenotes}[b]
	\item $^1$ Metallicities from \protect\cite{Walker:1992a,Grocholski:2006,Mucciarelli:2010,Mateluna:2012}, converted to CG97 metallicity scale where necessary (as in \protect\citealt{Wagner-Kaiser:2017}).
	\item $^2$ E(B--V) values from from Walker (1992, 1993), assuming R$_{V}$ =3.1.
	\item $^3$ Ages from \protect\cite{Wagner-Kaiser:2017}.
	\item $^4$ Magnitudes from \protect\cite{Mackey:2003a} (none for Reticulum).
	\item $^5$ Densities from \protect\cite{Mackey:2003a}.
\end{tablenotes}
\end{threeparttable}
\label{tab:LMC1}
\end{table*}

\begin{table*}
\caption{Fundamental properties of the LMC bar clusters in the \protect\cite{Olsen:1998} sample.}
\centering
\begin{threeparttable}[b]
    \begin{tabular}{@{}|l|c|c|c|c|c|c|@{}}
\textbf{Cluster} & \textbf{[Fe/H]$_{CG97}$$^1$} & \textbf{E(B--V)$^2$}  & \textbf{Age (Gyr)$^{3}$}   & \textbf{$\Delta$(V-I)} & \textbf{M$_V$$^{4}$} & \textbf{$\rho_0$$^{5}$}    \\  
\hline
NGC 1754 	& -1.30  	& 0.06	& 12.96 $\pm$ 2.2	& 0.750 $\pm$ 0.021	& 11.57	& 3.98	\\
NGC 1835 	& -1.79  	& 0.13	& 13.37 $\pm$ 2.8	& 0.696 $\pm$ 0.012	& 10.17	& 4.32	\\
NGC 1916 	& -1.54  	& 0.13	& 12.56 $\pm$ 5.5	& 0.807 $\pm$ 0.013	& 10.38	& 4.63	\\
NGC 2005 	& -1.54  	& 0.10	& 13.77 $\pm$ 4.9	& 0.865 $\pm$ 0.014	& 11.57	& 4.17	\\
NGC 2019 	& -1.67  	& 0.06	& 16.2 $\pm$ 3.1		& 0.825 $\pm$ 0.018	& 10.86	& 4.33	\\
\hline
    \end{tabular}
    \begin{tablenotes}[b]
	\item $^1$ Metallicities from \protect\cite{Olszewski:1991}, converted to CG97 metallicity scale.
	\item $^2$ E(B--V) values from \protect\cite{Walker:1992,Olszewski:1991,Olsen:1998,Johnson:2006,Pessev:2008}, assuming R$_{V}$ =3.1.
	\item $^3$ Ages from \protect\cite{Carretta:2010}, combined from \protect\cite{Olsen:1998} and \protect\cite{Beasley:2002}, and put on absolute scale assuming a reference of 13.5 Gyr (\protect\citealt{Marin-Franch:2009}).
	\item $^4$ Magnitudes from \protect\cite{Mackey:2003a}.
	\item $^5$ Densities from \protect\cite{Mackey:2003a}.
\end{tablenotes}
\end{threeparttable}
\label{tab:LMC2}
\end{table*}

Using the quantification of HB morphology with $\Delta$(V--I), we can compare to the values determined for the Galactic globular clusters from \cite{Dotter:2010}. However, differences in chemical abundances among clusters in each galaxy could make a direct comparison of HB morphology incomplete. Few studies have examined individual chemical abundances of individual stars in LMC clusters via high-resolution spectroscopy (\citealt{Hill:2000,Johnson:2006,Mucciarelli:2009,Mucciarelli:2010,Mateluna:2012}). Although some elements in the LMC cluster stars have been found to be distinct from GGC clusters (e.g.: [Cu/Fe], [Y/Fe]), thus far, the $\alpha$-abundances in the LMC globular clusters appear to fall in a range of values comparable to GGCs (\citealt{Johnson:2006,Mucciarelli:2010}), suggesting a direct comparison of $\Delta$(V--I) between the two galaxies is valid.

In Figure \ref{fig:dVIs}, the metal abundance [M/H] is calculated as in \cite{Salaris:1993}: [M/H] = [Fe/H]+$\log_{10}$(0.638 $\times$ 10[$\alpha$/Fe] + 0.362). We assume [$\alpha$/Fe] = 0.3 on average for the LMC clusters, as suggested by \cite{Mucciarelli:2009} for several of the clusters in our sample. We note that this assumption only affects the estimation of [M/H] as in Figures \ref{fig:dVIs} and \ref{fig:feh_resid}, and has little effect on our overall results. For [$\alpha$/Fe] values between 0 and 0.4, the values of [M/H] change minimally, by less than 0.2. For the GGCs, [$\alpha$/Fe] are taken from \cite{Dotter:2010} analysis for a direct comparison of their results to our own, though we note that \cite{Carretta:2010} and \cite{Nataf:2013} have additional [$\alpha$/Fe] estimates for the GGCs.

The LMC clusters are plotted as filled markers in Figure \ref{fig:dVIs}, with the red circles indicating the bar clusters from \cite{Olsen:1998} and the blue triangles representing the outer clusters from Paper I. The Galactic clusters are included with the same colour and shape convention as open markers, split into inner and outer halo clusters at R$_{GC}$ = 8 kpc. The dashed line in Figure \ref{fig:dVIs} is from \cite{Dotter:2010}, specifically their Table 2 and equations 2 through 4. This function was designed to represent the behavior of the inner halo clusters, whose HB morphology can be almost entirely characterized by their metallicities.

\subsection{Analysis}\label{HBanalysis}

In general, the LMC clusters - both the bar clusters and the outer clusters - largely occupy the same region in the metallicity-HB morphology space, having slightly redder HB morphologies than the older, inner Galactic GCs of comparable metallicity. There are no LMC clusters that have as blue HBs as the bluest Galactic clusters; it is possible this is due to differences in the helium abundances of the LMC and Galactic clusters. The primary outlier of the LMC clusters is Reticulum, whose HB is significantly redder than the other LMC clusters. However, it is worth noting that Reticulum has the most sparse HB of the clusters we analyze here, and the largest $\Delta$(V--I) error bar.

The bar clusters fall within 3 kpc of the LMC centre while the outer clusters are beyond $\gtrsim$ 4 kpc. However, there does not appear to be a clear differentiation in HB morphology between the LMC clusters located in the bar of the galaxy and those in the outer regions. The higher level of sensitivity at the blue end of the HB morphology spectrum with $\Delta$(V--I) allows us to see that as a population, the LMC clusters are more consistent with the outer halo Galactic population of clusters. While the inner halo clusters are scattered around the fitted line from \cite{Dotter:2010}, the LMC clusters do not follow suit. That the entire sample of LMC clusters are broadly consistent with the outer halo Galactic GCs lends further evidence to a Galactic formation scenario wherein the outer halo is built up from the accumulation of satellite galaxies similar to the LMC. However, it is also possible that this result is due to the restricted metallicity range of globular clusters in the LMC, in which there are no metal-rich clusters comparable to the Galactic population.

\begin{figure}
  \centering
    \includegraphics[width=0.5\textwidth]{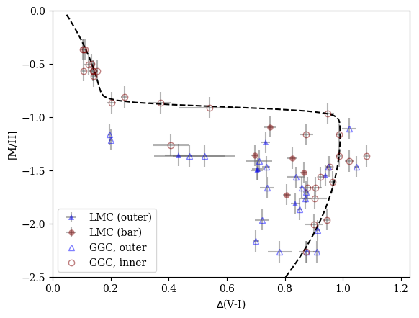}
  \caption{A comparison of the LMC clusters (solid markers) to the Galactic clusters (open markers). The inner clusters are shown as red circles for each galaxy and the outer clusters indicated as blue triangles. The HB morphology fitting functions from \protect\cite{Dotter:2010} are shown as the dashed line.}
  \label{fig:dVIs}
\end{figure}

\begin{figure}
  \centering
    \includegraphics[width=0.5\textwidth]{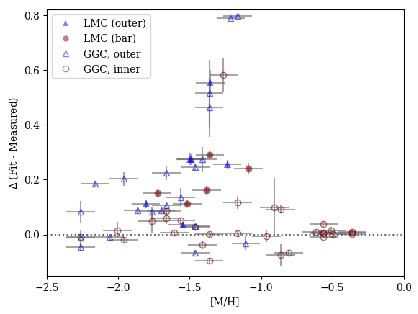}
  \caption{The residual between the equations from \protect\cite{Dotter:2010} to describe HB morphology and measured $\Delta$(V--I) values as a function of metallicity. The markers are the same as in Figure \ref{fig:dVIs}.}
  \label{fig:feh_resid}
\end{figure}

To examine the remaining deviation in HB morphology not explained by metallicity, we examine the residuals in $\Delta$(V--I) from Figure \ref{fig:dVIs}. This residual is calculated as the difference between the HB morphology predicted by the dashed line in Figure \ref{fig:dVIs} and the observed $\Delta$(V--I). In Figure \ref{fig:feh_resid}, this difference is plotted as a function of metal abundance. We see all the LMC clusters deviate from the expected $\Delta$(V--I) of the trend line, which is based only on metallicity (the first parameter). On average, the inner clusters deviate by 0.19 $\pm$ 0.07 (standard deviation) $\pm$ 0.01 (standard error) mag and the outer clusters deviate by 0.25 $\pm$ 0.16 (standard deviation) $\pm$ 0.03 (standard error) mag. Removing the influential point of Reticulum, the average deviation for the outer clusters is 0.19 $\pm$ 0.10 (standard deviation) $\pm$ 0.02 (standard error) mag, essentially equivalent to the inner clusters.
 
These results suggest broad similarity in HB morphology between the LMC clusters regardless of their physical location in the LMC. This is in stark contrast to the Milky Way, where the inner and outer halo clusters have markedly different HB morphologies.

\begin{figure*}
  \centering
    \includegraphics[width=\textwidth]{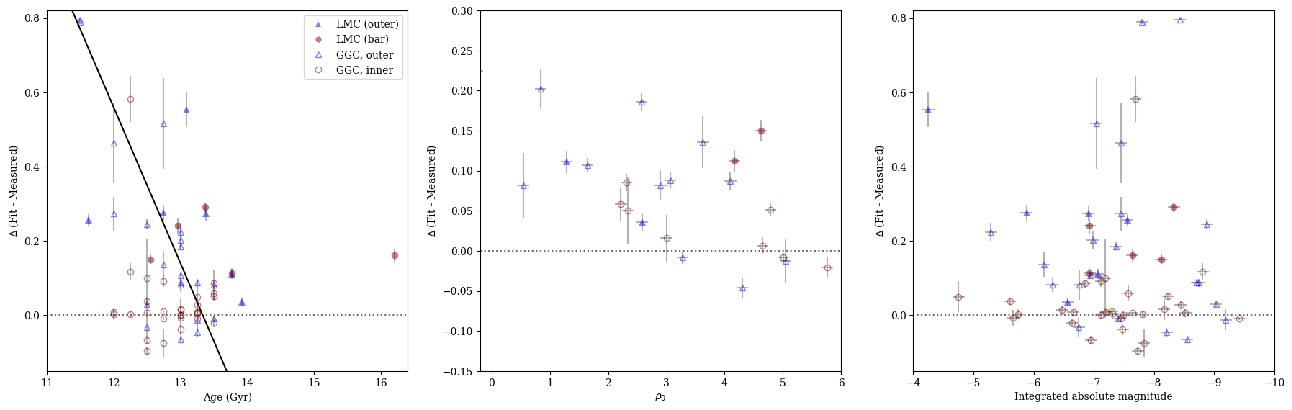}
  \caption{Left: the residual between the fitted and measured $\Delta$(V--I) values as a function of cluster age. The symbols are the same as in Figure \ref{fig:dVIs}. The bar cluster NGC 2019, at an estimated 16 Gyr, is the extreme outlier. The solid line is the fit from \protect\cite{Dotter:2010}. Middle: $\Delta$(V--I) residual compared to cluster central densities (\protect\citealt{Mackey:2003a}). As in \protect\cite{Dotter:2010}, we plot only the metal-poor clusters with [M/H] $\textless$ --1.5. Right: residuals compared to cluster integrated magnitudes (\protect\citealt{Geisler:1997,Mackey:2003a}).}
  \label{fig:all_resid}
\end{figure*}

As to what else may contribute to HB morphology beyond metallicity - the possible second parameter(s) - we further explore the deviation of clusters from the HB morphology expected from metallicity alone. We plot the residuals in $\Delta$(V--I) with age, cluster central density, and integrated magnitude in Figure \ref{fig:all_resid}, with Milky Way values from \cite{Dotter:2010}. LMC cluster ages are from \cite{Wagner-Kaiser:2017} for the outer clusters and \cite{Beasley:2002,Olsen:1998} for the inner bar clusters (as presented in \citealt{Carretta:2010}). Densities are taken from \cite{Mackey:2003a} and integrated magnitudes from \cite{Geisler:1997}. Table \ref{tab:spearman} provides the correlations for these relations and that of [M/H] (as in Figure \ref{fig:feh_resid}) for the LMC and Galactic globular clusters combined.

With age, the leftmost panel of Figure \ref{fig:all_resid}, our results are consistent with that of \cite{Dotter:2010}. The outer clusters show a statistically significant trend between age and the residual of $\Delta$(V--I) with a Spearman $\rho$ of --0.50 (p-value = 0.008). This is qualitatively similar to the results for the outer clusters from \cite{Dotter:2010}, with a Spearman $\rho$ of --0.81. Our results are also in agreement with the findings of \cite{Milone:2014}, who also demonstrate a statistically significant relationship between HB morphology and age (albeit using a different method of quantifying the structure of the HB). 

For the LMC bar clusters, however, we do not observe any statistically significant relationship between age and the $\Delta$(V--I) residual, again consistent with the results from \cite{Dotter:2010}. While there is no additional constraining information beyond what is already known from the Galactic population, the LMC clusters are consistent with the Galactic trends. We do note that the oldest cluster, NGC 2019, is a significant outlier at 16 Gyr; however, the quoted uncertainties in \cite{Olsen:1998} and \cite{Beasley:2002} are on the order of several Gyr.

In examining the central densities of the clusters, a similar pattern emerges. The relation with central density for clusters with [M/H] $\textless$ --1.5 is fairly convincing, and the LMC clusters are again consistent with the trend seen in the Galactic clusters. The correlation is statistically significant with a Spearman correlation of --0.41.


For the absolute integrated magnitude, the LMC clusters present similar scatter as seen in the Milky Way clusters, though the data show no clear trend. The relationship has a Spearman correlation coefficient of 0.10, consistent with the results of \cite{Dotter:2010}, who find a Spearman correlation coefficient of 0.11. However, \citealt{Milone:2014} have showed that their L2 parameter, which measures the overall colour extension of the HB, correlates with the absolute magnitude of the clusters. This finding implies that the absolute integrated cluster magnitude correlates with helium variations, with larger helium variations in more luminous (massive) clusters. Although we do not see such a correlation, this does not mean there is no such correlation, as our choice of HB parameter is not sensitive to helium variations, as previously discussed.

\begin{table}
\caption{Spearman Correlation Results for Figures \ref{fig:feh_resid} and \ref{fig:all_resid}.}
\centering
    \begin{tabular}{@{}|l|c|c|@{}}
\textbf{Variable} & \textbf{$\rho$} & \textbf{p-value} \\  \hline
$[$M/H$]$&	-0.08		&	0.51		\\
Age (Inner)&	0.15		&	0.38		\\
Age (Outer)&	-0.50		&	$\textless$0.01  \\
Density	&	-0.41		&	0.02		\\
M$_{V}$	&	0.09		&	0.46		\\
\hline
    \end{tabular}
\label{tab:spearman}
\end{table}

Hodge 11 exhibits a bluer horizontal branch than the other LMC clusters we examine. If this extension is driven by helium, we expect that future observations of Hodge 11 should demonstrate that the cluster has the largest internal helium variation amongst the clusters considered here. However, we note that Hodge 11 does not appear to be the brightest (or most massive) cluster of this cluster sample, which does not seem consistent with what would be expected. At present there is little or no quantitative assessments of these parameters for the LMC clusters. Our understanding of these clusters would benefit from further work exploring these possibilities.

NGC 2210 is another intriguing cluster in the LMC sample. This cluster is thought to be about 1.5 Gyr younger than the other LMC clusters (\citealt{Wagner-Kaiser:2017}). However, its median HB colour does not reflect this, even though $\Delta$(V--I) is age-sensitive. Although NGC 2210 has a bluer HB than predicted, it does not have an unusually strong blue HB extension. If blue HB extensions are driven by helium enhancements, then helium may not be the cause of the HB variation in NGC 2210. It may be that the higher central density (compared to the other outer LMC clusters) could be the cause of the bluer than expected HB. The same may be true for Reticulum, which has the reddest median HB colour of the clusters we examine, yet does not have a markedly different age or metallicity from the other LMC clusters.



\section{Conclusions}\label{conclusions}

In this paper, we have examined the HB morphology of globular clusters in the Large Magellanic Cloud and compared their characteristics to the Milky Way globular clusters. Our findings  include:

\begin{enumerate}  

\item The inner (bar) and outer clusters in the LMC are generally comparable in their HB morphology characteristics as described by the $\Delta$(V--I) measure. Despite their galactocentric differences, they occupy the same region in metallicity-HB morphology space.

\item As observed for the outer halo population of GCs, the LMC clusters deviate from HB morphologies that can be explained by metallicity alone. This is a clear indication that these clusters exhibit the classical second parameter effect, which is consistent with the idea that the outer halo Galactic population was accreted from dwarf satellites as originally advocated by Searle \& Zinn (1978).

\item Despite the fact that the HB morphologies of the LMC clusters are clearly affected by at least one parameter in addition to metallicity, we do not find any convincing evidence to uniquely identify this parameter or parameters. The LMC clusters generally agree with the trends seen in Galactic GCs between median HB colour and age, central density, and integrated absolute magnitude. Reticulum, NGC 2210, and Hodge 11 constitute good examples of clusters for which age cannot be the sole second parameter. Internal helium variations could be important for Hodge 11, which has a strongly-extended HB, however this does not appear to be the case for Reticulum or NGC 2210.

\end{enumerate}


\section*{Acknowledgments}
A.D.M. is grateful for support from an Australian Research Council (ARC) Future Fellowship (FT160100206). D.G. gratefully acknowledges support from the Chilean BASAL Centro de Excelencia en Astrof\'isica y Tecnolog\'ias Afines (CATA) grant PFB-06/2007. We thank an anonymous referee whose comments and suggestions were very helpful.

\bibliographystyle{mn2e}
\bibliography{LMC}
\clearpage

\label{lastpage}

\end{document}